\documentclass[aps,prl,nofootinbib,floatfix,preprintnumbers,twocolumn,amsmath,amssymb]{revtex4}
\usepackage{graphicx}
\usepackage{pdfpages}
\usepackage{multirow}
\usepackage[utf8]{inputenc}
\usepackage[english]{babel}
\usepackage{amsmath}
\usepackage{graphicx}
\usepackage{color}
\usepackage{amssymb}
\usepackage{hyperref}
\usepackage{dcolumn}% Align table columns on decimal point
\usepackage{bm}% bold math

\def\beq{\begin{eqnarray}}
\def\eeq{\end{eqnarray}}

\begin{document}

\title{A new radiative neutrino mass generation mechanism with higher dimensional scalar representations and custodial symmetry.}

\author{
Alfredo Aranda,$^{a,b}$\footnote{Electronic address: fefo@ucol.mx}
Eduardo Peinado,$^{c}$ \footnote{Electronic address: epeinado@fisica.unam.mx}
\vspace*{0.3cm}}

\affiliation{$^{(a)}$ Facultad de Ciencias, CUICBAS, Universidad de Colima, 28040 Colima, Mexico;}
\affiliation{$^{(b)}$ Dual C-P Institute of High Energy Physics, 28040 Colima, Mexico;}
\affiliation{$^{(c)}$ Instituto de Física, Universidad Nacional Autónoma de México, A.P. 20-364, M\'exico D.F. 01000, M\'exico.}

\date{\today}

\begin{abstract}
A new realization for radiative neutrino mass generation is presented. Based on the requirement of tree-level custodial symmetry and the introduction of higher (greater than two) dimensional representations for scalar fields, a specific scenario with a scalar septet is presented that generates neutrino Majorana masses radiatively. This is accomplished through an eleven dimensional operator that requires the addition of several scalar fields and a SU(2) 5-plet of new fermions, together with a $Z_2$ that guarantees the preservation of custodial symmetry. The phenomenology of the setup is rather rich and includes a dark matter candidate.
\end{abstract}
\maketitle
 
Precision electroweak measurements constitute a very clear window into possible new physics effects~\cite{Agashe:2014kda}. In particular, the {\it observable} related to the custodial symmetry present in the scalar sector of the Standard Model (SM) called $\rho$~\cite{Ross:1975fq,Veltman:1977kh,Einhorn:1981cy}, puts stringent constraints on any scalar extension related to electroweak symmetry breaking (EWSB). The SM prediction for this parameter is $\rho = 1$, in complete agreement with measurement~\cite{Agashe:2014kda}. 

An inescapable fact is that extending the scalar sector of the SM is practically a must in any scenario of physics beyond it. In particular, the addition of extra SU(2) doublet scalars has led to a rich source of phenomenological studies with a vast literature, see for instance \cite{Lee:1973iz,Deshpande:1977rw}. A key feature singled out in this letter is the fact that scalar SU(2) doublets do maintain the custodial symmetry present in the SM and hence the prediction of $\rho=1$ at tree level in agreement with measurement, but that in order to use other representations one must go to high values such as a septet in the next minimal case (triplets can be used with specific conditions but are strongly constrained\cite{Georgi:1985nv,Chanowitz:1985ug,Lynn:1990zk,Hung:2006ap,Aranda:2008ab}).

Seemingly unrelated to the issue of custodial symmetry, but not to the scalar sector of the SM and models beyond it, is the problem of neutrino masses. Stated succinctly: the generation of neutrino masses generically {\it demands} some extended scalar sector. A very important and well known feature of neutrinos is that there are two options for their fermionic nature: they can be either Dirac particles, like the rest of the fermions in the SM, or they can be Majorana. Models with Majorana neutrino masses abound that rely on basically two general approaches (sometimes intertwined): radiative mass generation~\cite{Zee:1980ai,Cheng:1980qt,Zee:1985id,Babu:1988ki,Pilaftsis:1991ug}\footnote{For a systematic clasification of one-loop and two-loops realizations of the dimension 5 Weinberg operator see~\cite{Bonnet:2012kz} and~\cite{Sierra:2014rxa}.}, and the so-called {\it seesaw mechanism} \cite{Minkowski:1977sc,Yanagida:1980xy,GellMann:1980vs,Ramond:1979py,Mohapatra:1979ia,Mohapatra:1980yp,Schechter:1980gr,Schechter:1981cv,Magg:1980ut,Lazarides:1980nt,Foot:1988aq} (in its several implementations and schemes).\footnote{In supersymmetric extensions with R-parity violation it is possible to account for Majorana neutrino masses~\cite{Aulakh:1982yn,Hall:1983id,Ross:1984yg}} For detailed descriptions and comprehensive references on these please see~\cite{Mohapatra:1998rq,Valle:2015pba}. Scenarios with Dirac neutrinos, although less popular, also exist. Some of them use, for example, extra dimensional constructions~\cite{ArkaniHamed:1998vp,Hung:2002qp,Gherghetta:2003hf}, seesaw-type mechanisms~\cite{Farzan:2012sa,Ma:2014qra,Ma:2015raa}, extra Higgses that couple only to neutrinos~\cite{Wang:2006jy,Gabriel:2006ns,Davidson:2009ha,Memenga:2013vc}, flavor symmetries that forbid the existence of Majorana fermions~\cite{Aranda:2013gga}, and combinations thereof~\cite{Ding:2013eca}. It is worth to mention that higher SU(2) representations have been used to generate neutrino masses, see for instance~\cite{McDonald:2013kca,Law:2013gma,Culjak:2015qja,Ahriche:2015wha}. For an interesting recent proposal using new gauge interactions see~\cite{deGouvea:2015pea}.

In this letter a new scenario is presented in the context of radiative Majorana neutrino mass generation that  is motivated and restricted by custodial symmetry. By imposing as a requirement an extended scalar sector that preserves custodial symmetry, the question then is what kind of minimal construction can be proposed in order to successfully generate Majorana neutrino masses.

It is well known that, barring SU(2) doublets, custodial symmetry preservation (at tree level) requires {\it minimally} a SU(2) septet with Hypercharge $Y=2$~\cite{Gunion:1989we}. The scalar potential of such a septet has been analyzed in detail in for example~\cite{Hisano:2013sn}. A crucial observation is the existence of a dimension seven operator that induces a vacuum expectation value (vev) of the septet and guarantees a proper electroweak vacuum (the presence of the dimension seven operator guarantees the absence of an exact massless Nambu-Goldston boson). For a thorough discussion on constraints to septet-doublet mixing from precision electroweak parameters see~\cite{Geng:2014oea}.

The model in this paper takes those initial ingredients (SM particle content plus a scalar SU(2) septet) and adds the minimal structure required to obtain a (radiative) mechanism for neutrino masses. The complete particle content is then that of the SM plus the following fields (multiplets referring to SU(2)): a fermionic quintet $\Psi$ with Hypercharge $Y_f=0$, a septet scalar $\chi$ with $Y_7=2$, a scalar quartet $\Theta$ with $Y_4=1/2$, and a quintet scalar $\eta$ with $Y_5=-1$. 

In order to preserve the custodial symmetry, the $\eta$ and $\Theta$ scalar fields must have zero vevs. This is guaranteed by means of a $Z_2$ symmetry. Furthermore, in order to successfully induce the Majorana mass for the light neutrinos, the fermionic quintet $\Psi$ must also be charged under the $Z_2$ symmetry. As a consequence the scenario can lead to either scalar or fermionic candidates for dark matter.  This is the minimal matter content for a framework where the septet contributes to Majorana Neutrino masses. To see that, consider that the fermion (with hypercharge $Y_f=0$) must contain a neutral component and thus must be in an odd SU(2) representation. If assigned to a triplet, it couples directly to the SM Higgs leading to the Type III seesaw~\cite{Foot:1988aq} and thus the septet is {\it unnecessary}. The quintet is therefore the smallest representation that works for the mechanism presented in this work (the additional scalar four and five dimensional fields correspond to the minimal set required in order to {\it close the loop}).

The vev of $\chi$ is induced through the dimension seven operator required for the proper electromagnetic vacuum mentioned before~\cite{Hisano:2013sn} (note that the $Z_2$ symmetry prevents $\Theta$ and $\eta$ from forming similar vev inducing operators): 
\begin{eqnarray}\label{dim7operator}
{\cal L}\supset -\frac{1}{\Lambda^3} \left( \chi^{abcdef}\phi_a\phi_b\phi_c\phi_d\phi_e\phi^{*g}\epsilon_{fg} + h.c. \right) ,
\end{eqnarray}
with $\phi_1$ and $\phi_2$ denoting the charged and neutral components of the SM SU(2) doublet respectively: $\phi_1 = \phi^+$ and $\phi_2 = (\phi_0 + v +iA)/\sqrt{2}$. The (symmetric) tensor notation used for the septet $\chi$ with components $\chi_i$ ($i$ denoting the isospin component) is $\chi_{111111} = \chi_{3}$, $\chi_{111112} = \chi_{2}/\sqrt{6}$, $\chi_{111122} = \chi_{1}/\sqrt{15}$, $\chi_{111222} =\chi_{0}/\sqrt{20}$, $\chi_{112222} = \chi_{-1}/\sqrt{15}$, and $\chi_{122222} = \chi_{-2}/\sqrt{6}$, and $\chi_{222222}=\chi_{-3}$ with $\chi_{-2} = (v_7+h_7+iA_7)/\sqrt{2}$.

\begin{figure}[ht]
\includegraphics[scale=0.3]{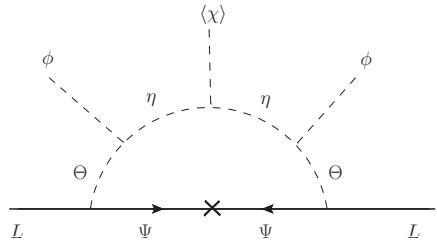}
\caption{Diagram for neutrino mass generation involving, in addition to the usual SM fields for left-handed Majorana neutrino masses, the fermionic SU(2) 5-plet $\Psi$, the vacuum expectation value of the scalar septet $\chi$ associated to custodial symmetry, and the additional required scalars in the quartet ($\Theta$) and quintet ($\eta$) SU(2) representations.}
\label{figure1}
\end{figure}

A Lepton number violating Majorana mass for the neutral component of $\Psi$ is present and required for the radiative mechanism that induces small Majorana neutrino masses. This can be seen explicitly in the diagram shown in Figure~\ref{figure1}. Note that such a diagram corresponds to an effective dimension eleven Weinberg-like operator $LL\phi\phi\widetilde{\chi}$, where $\widetilde{\chi}$ denotes the effective dimension six operator implicit in the diagram.
%\acknowledgments
\vspace{0.5cm}

Several observations regarding the setup can be singled out: the complete scenario is motivated by an extension that respects custodial symmetry. Hence, once scalar SU(2) representations other than doublets are considered, the septet is the minimal choice. This determined, the introduction of the fermionic quintet represents the minimal setting for the generation of neutrino masses in a novel fashion. This $7 - 5$ / scalar - fermion combination is therefore unique and justifiable.

To accomplish the radiative generation of mass more is needed. The introduction of  additional scalars is necessary and the most economical case corresponds to the inclusion of $\Theta$ and $\eta$ in such a way that they do not participate in EWSB: an additional $Z_2$ symmetry is incorporated to accomplish that. Note that the neutrino mass scale is given by the combined suppressions of loop factors and the (perhaps not so heavy) scalars running in the loop. Although a complete computation for the neutrino mass is out of the scope of this letter (and will be presented elsewhere together with a comprehensive analysis of the full scalar sector), it is possible to make a general estimate to determine the necessary scales.
Assuming all dimensionless couplings to be in the range of O$(10^{-1}) - \rm{O}(1)$, and taking the characteristic (dominant) mass scale running in the loop to be of O$(10 ~\rm{TeV})$
an acceptable neutrinos mass scale is obtained for a septet vev in the range of O$(10^{-6} - 10^{-1}~\textrm{GeV})$. 

All these ingredients lead to a rich scalar and fermionic phenomenology potentially testable at colliders: this scenario provides a {\it low scale Majorana neutrino mass mechanism}.

The presence of the $Z_2$ symmetry  that guarantees custodial symmetry also gives the possibility for a stable neutral field coming from either a combination of the neutral components of $\Theta$ and $\eta$, or from the neutral component of $\Psi$, and so the framework also provides a dark matter candidate. If it comes from the scalar sector it is a so-called {\it Higgs portal} scenario where the coupling present in neutrino mass generation is also present in dark matter direct detection (and also probably its relic density), as is typical in other known cases such as, for example, the scotogenic model~\cite{Ma:2006km}. What is peculiar and interesting about the present setting is the connection between custodial symmetry preservation and the possibility of scalar dark matter.

If the dark matter candidate is fermionic and thus comes from the quintet $\Psi$, it is also important to note that it corresponds to the preferred case (for O(TeV) mass) of the so-called minimal dark matter scenario analyzed in~\cite{Cirelli:2005uq}, for a more recent analysis on this topic, see~\cite{DiLuzio:2015oha,Hamada:2015bra}. This is an interesting and noteworthy coincidence since the quintet representation has been required/motivated in this scenario by merging custodial symmetry with radiative neutrino mass generation, a totally (a priori) unrelated sector.

A potential concern with frameworks that use high dimensional representations is that of perturbativity: the use of higher dimensional representations alter the running of gauge couplings in such a way that they might become non-perturbative very quickly. An analysis based on the requirement of perturbativity for the SU(2) gauge coupling was performed in~\cite{Cirelli:2005uq} leading to the following upper bounds on the dimension ($n$) for viable representations: $n \leq 5$ for Majorana fermions and $n \leq 8$ for scalars. The scenario presented here satisfies both bounds.

The phenomenology derived from this new scenario for radiative neutrino mass generation, together with the two interesting possibilities for scalar or fermionic dark matter\footnote{For an estimate of the relic density of this two scenarios, see~\cite{Cirelli:2005uq}} (considered to be an important spinoff of this scenario) deserve further study and are currently under investigation.

{\bf Acknowledgements}
E.P. thanks Universidad de Colima for its hospitality while part of this work was being carried out. This work has been supported in part by PAPIIT IA101516, PAPIIT IN111115, SNI, CONACYT 132059 and CONACYT CB-2011-01-167425 (Fondo Sectorial de Investigaci\'on para la Educaci\'on - M\'exico).

\bibliographystyle{ieeetr}

\end{document}